\documentclass[utf8]{frontiersSCNS} 

\usepackage{url,hyperref,lineno,microtype,subcaption}
\usepackage[onehalfspacing]{setspace}
\usepackage{enotez}

\DeclareInstance{enotez-list}{custom}{list}
{
    heading = \section*{Footnotes},
    list-type = itemize,
    format = \normalfont,
    number = \textsuperscript{#1},
}

\def\keyFont{\fontsize{8}{11}\helveticabold }
\def\firstAuthorLast{Kim WJ, Khomtchouk BB, 2024} 
\def\Authors{Woo Jae Kim\,$^{1,2}$, Bohdan B. Khomtchouk\,$^{2}$}
\def\Address{
  $^{1}$The College of the University of Chicago, Chicago, IL USA \\
  $^{2}$Department of Biomedical Engineering \& Informatics, Luddy School of Informatics, Computing, and Engineering, Indiana University Indianapolis, Indianapolis, IN USA
}
\def\corrAuthor{Bohdan B. Khomtchouk}
\def\corrEmail{bokhomt@iu.edu}

\begin{document}
\onecolumn
\firstpage{1}

\title[Wasm enables low latency interoperable AR/VR software]{WebAssembly enables low latency interoperable augmented and virtual reality software} 

\author[\firstAuthorLast ]{\Authors} 
\address{} 
\correspondance{} 

\extraAuth{}

\maketitle

\begin{abstract}

There is a clear difference in runtime performance between native applications that use augmented/virtual reality (AR/VR) device-specific hardware and comparable web-based implementations. Here we show that WebAssembly (Wasm) offers a promising developer solution that can bring near-native low latency performance to web-based applications, enabling hardware-agnostic interoperability at scale through portable bytecode that runs on any WiFi or cellular data network-enabled AR/VR device. Many software application areas have begun to realize Wasm’s potential as a key enabling technology, but it has yet to establish a robust presence in the AR/VR domain. When considering the limitations of current web-based AR/VR development technologies such as WebXR, which provides an existing application programming interface (API) that enables AR/VR capabilities for web-based programs, Wasm can resolve critical issues faced with just-in-time (JIT) compilation, slow run-times, large file sizes and big data, among other challenges. Existing applications using Wasm-based WebXR are sparse but growing, and the potential for porting native applications to use this emerging framework will benefit the web-based AR/VR application space and bring it closer to its native counterparts in terms of performance. Taken together, this kind of standardized ``write-once-deploy-everywhere" software framework for AR/VR applications has the potential to consolidate user experiences across different head-mounted displays and other embedded devices to ultimately create an interoperable AR/VR ecosystem (\cite{jacobsson}).

\tiny
 \keyFont{ \section{Keywords:} Augmented Reality, Virtual Reality, WebAssembly, Software Engineering, Performance, Interoperability} 
\end{abstract}

\section{Introduction}

From a consumer's point of view, the range and variety of augmented and virtual reality (AR/VR) software experiences available on any given hardware device will ultimately drive user buying preferences and long-term retention.  Therefore, software developers are well-advised to create codebases that are future-proofed to run on as many internet-connected AR/VR devices as possible, bringing together user experiences regardless of hardware vendor preferences. As immersive technologies such as AR/VR, jointly referred to as extended reality (XR), continue to rapidly develop in their scope and availability, the limitations of current strategies in computation and adoption are becoming increasingly more apparent. Currently available commercial headsets are almost exclusively dedicated to native-run applications with few offering AR support. Mobile phones, which offer the greatest opportunity for portable and accessible XR, are unfortunately limited by computational power and a lack of development focus\footnote{https://www.androidauthority.com/challenges-facing-mobile-vr-771609/}. Mobile VR headsets such as Google Cardboard and Samsung Gear VR, in which the user inserts a mobile phone into a headset to serve as the display, have faded among both developers and users who favor more dedicated solutions such as the Oculus or Vive headsets. Meanwhile, AR on mobile phones has seen incremental improvements in general, such as the continued optimization and functionality of the ARCore and ARKit APIs from Google and Apple, respectively (\cite{nowacki}). Despite this, there is still a lack of demand for mobile AR applications or a standardized marketplace that consolidates them together within an interoperable ecosystem (\cite{braud}).

The closest solution to the accessibility bottleneck in the XR arena is the WebXR API, a set of open-source standards produced collaboratively by companies including Google and Mozilla, designed to support cross-platform web-based XR experiences\footnote{https://github.com/immersive-web/webxr/blob/master/explainer.md}.\label{en:webxr} While WebXR offers a promising revitalization of a web-based XR experience, it is still incompatible with most major browsers and has significant performance limitations. With XR being one of the most resource-intensive activities intended for common use, current web-based implementations cannot match the efficiency of native XR. Hence, there are few multi-purpose WebXR tools beyond a simple demo or proof of concept (\cite{renius}).

To accelerate the adoption and utility of web-based cross-platform XR experiences, we propose WebAssembly (Wasm) as a solution for computationally expensive XR operations. Wasm is a low-level binary format that markedly improves application performance to near-native speeds (\cite{perkel}). Another advantage is that it is both a compilation target for a range of languages like C++, C\#, or Rust, and completely platform-agnostic as it runs fully on the web (\cite{bbb, alamari}). We propose a shift to Wasm for every major step involved in XR visualization: tracking, rendering, registration, data visualization, and more. By reaching near-native speeds, Wasm with WebXR has the possibility to combine the accessibility and cross-platform advantages of WebXR with the computational advantages of native solutions. In addition, once Wasm gets sufficient open-source XR attention and traction among developers, existing native XR applications can be ported over to the web for future maintenance and deployment.

\section{Web-based AR/VR}
\subsection{WebVR/XR API}

As compilers, languages, and APIs grow more efficient, web substitutes of native applications become more capable and widespread. One notable example is Microsoft Office Web released in 2010, which provides access to Microsoft Office applications through a web browser. As more complex web applications became more common in the early-mid 2010s, aided by the high-profile releases of HTML5 (with Web Workers API) and WebGL, web-based VR also started appearing in simple demos (\cite{qiao}). Mozilla commenced development of the original WebVR API in 2014 and finally released the first version in early 2016. The latest iteration of this project, the WebXR standard, was released in 2018 and is actively maintained to this day.

From its inception, WebVR’s primary goal was to standardize web-based VR experiences by querying XR capabilities, recognizing controller inputs, and displaying images on the headset. Hence, the WebVR API provides access to the input and output but does not handle the rendering of the 3D graphics itself. Instead, this operation was delegated to WebGL, a graphics API for web browsers (\cite{macintyre}). There are many frameworks such as Mozilla’s A-Frame\footnote{https://aframe.io} and Facebook’s React 360\footnote{https://github.com/facebookarchive/react-360} built to supplement WebVR in an effort to further simplify web-based VR development. These frameworks use Javascript (JS) 3D libraries such as Three.js for the rendering step. 

Beyond the technical implementation, the early stages of WebVR signaled a greater shift within the history of high-performance web-based applications. While the early era of WebVR experiences mostly consisted of still 3D images rendering at 20 frames per second (\cite{hubbold}), it still laid the groundwork and standardization protocols for more efficient cross-platform XR experiences. Compared to WebVR, WebXR featured notable changes in implementation and augmented reality support \endnotemark[\ref{en:webxr}]. While the scalability and implementation of WebXR underwent a major overhaul, the goals and vision of the project still remained similar\footnote{https://developer.mozilla.org/en-US/docs/Web/API/WebXR\_Device\_API/Fundamentals}.

While the release and adoption of WebXR has been successful -- with popular cross-platform experiences used commonly across desktop computers, VR headsets, and mobile phones -- it still has a relatively small commercial presence compared to native alternatives. Even with its current maturity, there still only exist a handful of fully-featured applications. An example is Mozilla Hubs, a collaborative meeting space in VR (\cite{macintyre}). However, most applications using WebXR so far are mere proof-of-concept implementations (\cite{renius}).

Unfortunately, WebXR’s small user base is due to many self-perpetuating factors. While WebXR offers an exciting cross-platform solution, it is still incompatible with most major web browsers, especially amongst mobile phones (\cite{qiao, renius}). Currently, only Chrome, Microsoft Edge, and Firefox are the major browsers that support WebXR. The only way to run WebXR on iOS is to use the WebXR Viewer app, which essentially defeats the purpose of a cross-platform browser-based solution. With few devices supporting WebXR, there is less demand for a fully integrated user experience. And given the lack of development and commercialization, there is less demand for platform adoption. On virtual reality headsets, where WebXR is currently supported and performs well, native application marketplaces offer a better variety of experiences, ease of discovery, and offer financial incentives for developers.

\subsection{Browser Compatibility}

In spite of the drawbacks, the advantages of web-based AR/VR are clear -- for example, the ease and simplicity for users to share and load content, interconnectedness of the web, among many other factors (\cite{zhangYaping}). However, as noted in the previous section, a key limitation of web-based XR frameworks is cross-platform compatibility (\cite{macintyre}). Given the current lack of a standardized framework, universal compatibility across all major browsers is not a reality as of yet, with WebXR coming the closest. Access to all platforms including all major browsers remains an issue, though Mozilla has committed to expanding WebXR’s cross-platform compatibility.

\subsection{Performance and Latency}

Another major disadvantage is the significant impact on the performance and latency of web applications in general (\cite{yan}). Most modern JS uses Just-In-Time compilation (JIT), which compiles the code during run-time. While optimization in compilers has drastically improved, JIT lags behind Ahead-Of-Time compilation (AOT), which converts the code during the build-time before being run. The difference is due to a shorter compilation time constraint, which offers less flexibility for low-level optimization. Furthermore, there are occurrences of re-optimization where parts of the optimized code are discarded (due to incorrect assumptions made by JIT) and then sent through the optimizing compiler again. This process of re-optimizing is significantly time-consuming. In addition, JS is particularly prone to “JIT-unfriendly” code, as it contains many popular dynamic features that are used inconsistently (\cite{gong}). One other large factor is the fetching speed of a JS bundle, where the transfer time from the server to the client is not optimal even after being compacted, primarily over slow networks (\cite{jiang}). Even an application that may be optimized for performance, containing large amounts of visual and audio assets, will be much more difficult to deliver instantaneously compared to an offline application. 

Altogether, the advances in WebXR standards and experiences have been impressive, but still not fully utilized. While adoption is currently slow, iterative improvements in performance efficiency, browser compatibility, and developer attention will eventually enable WebXR to cross a threshold into the mainstream. Of all these limitations, we believe that near-native performance and platform compatibility offer the greatest opportunity for WebAssembly to accelerate widespread adoption.

\section{WebAssembly}

\subsection{History and Design}

WebAssembly (Wasm) is a young, low-level binary format that brings near-native speeds to the web. It can be a compilation target for a variety of languages, including typed languages such as C/C++ among many others, including Python\footnote{https://github.com/pyodide/pyodide}. Thus, Wasm opens the door to porting many existing native applications to the web\footnote{https://webassembly.github.io/spec/core/}.\label{en:wasm}

The precursor to Wasm was asm.js, a language also intended as a compilation target, specifically from JS code (\cite{murphy}). Its main performance optimizations come from its usage of AOT compilation. However, due to its inherent design as a JS compilation target, asm.js inherits many of the same limitations from JS, such as relatively slow compilation times. Additionally, the generated asm.js text is more difficult to parse than binaries. Thus, Wasm was created as a successor, which would compile native code into binaries, avoiding the main limitations of asm.js while retaining the many optimizations such as AOT compilation. Thus, Wasm offers considerable improvements over asm.js, as it can be a compilation target for a multitude of languages, not just JS. This subsequently provides access to other source codes, libraries, etc. that are available through the web.

\begin{figure}[h!]
\begin{center}
\includegraphics[width=10cm]{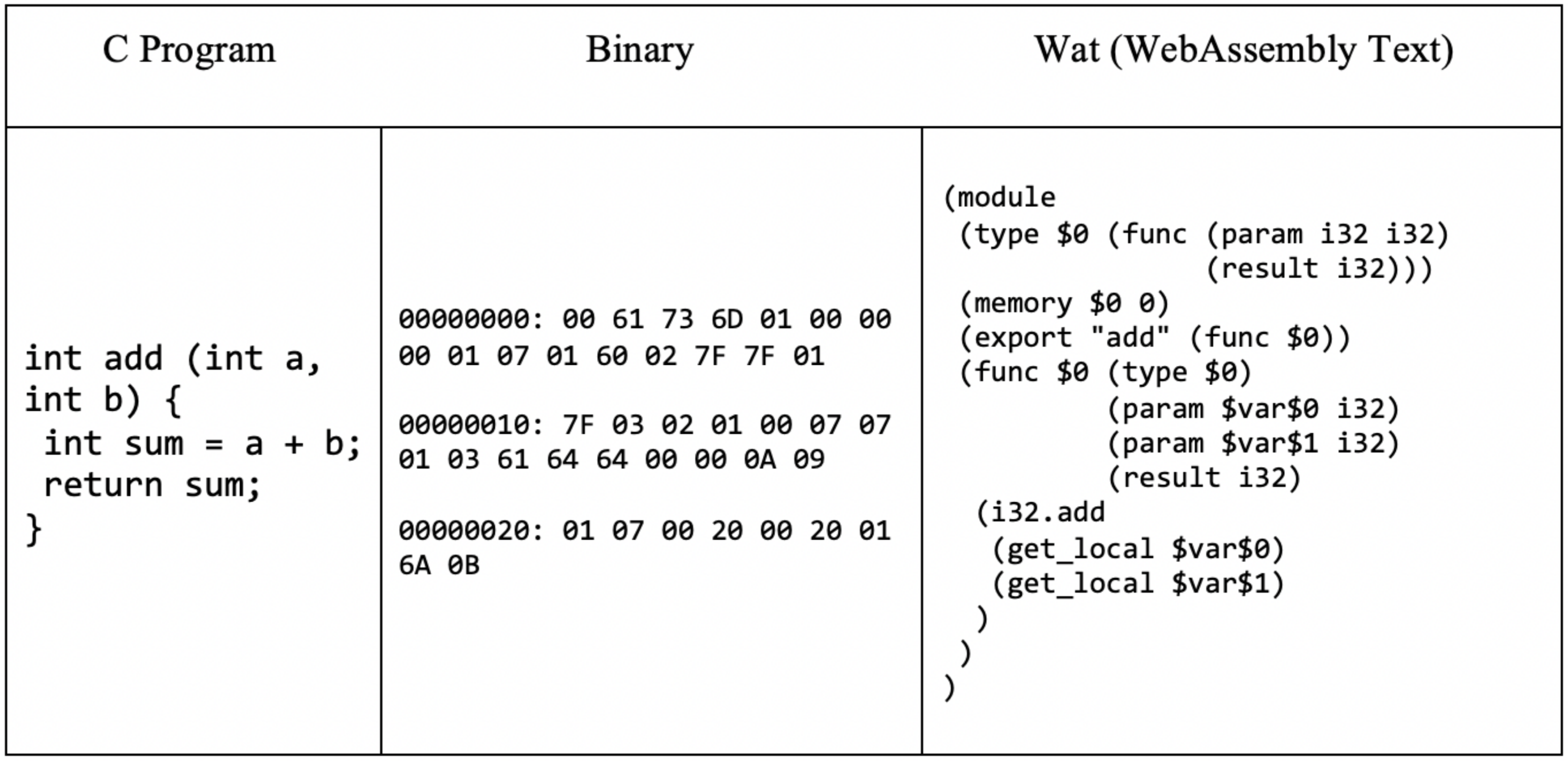}
\end{center}
\caption{Example of a simple native C program of addition that is compiled into a binary format using WebAssembly}\label{fig:1}
\end{figure}

Wasm modules can be embedded into separate execution environments, and have primarily been used with JS and Web APIs to run in the browser (\cite{rossberg}). It is important to note that Wasm was designed to work in tandem with JS as opposed to being its replacement. Wasm's main advantages are that it can bring source language code to the web, can communicate effectively with JS, and can run at higher speeds and smaller sizes (\cite{kjorveziroski}). This performance advantage stems from its comparatively faster fetching speed than JS, more efficient instruction parsing, AOT compilation, and manual memory management that removes the need for garbage collection\cite{jangda}\footnote{https://hacks.mozilla.org/2017/02/what-makes-webassembly-fast/}. found that most benchmarks came within 10\% of native code, and code size was 63\% the size of equivalent JS, and 85\% the size of native code. On analyzing the differences between platforms, \cite{rossberg} found that Wasm even outperforms some native benchmarks and significantly outperforms JS (by 150-250\%, depending on the platform). As Wasm and its implementation continue to evolve, we expect the performance gap to widen in Wasm’s favor. Since its minimum viable product release in 2017, major web browsers rapidly saw the utility of Wasm and began supporting it. As of now, over 93\% of browsers support it. In addition, many popular languages now have compilers and frameworks for constructing Wasm, representing a huge leap forward in innovation compared to its precursor asm.js.

\subsection{WebAssembly Implementation}

The first major languages to support Wasm were C/C++ through Emscripten, a compiler toolchain to port LLVM bitcode to Wasm. It also converts OpenGL, a graphics library designed for installed software, to WebGL, which is its web-based counterpart. Besides creating the Wasm module, Emscripten generates the JS and HTML necessary for loading on the web, allowing for web applications to easily integrate into existing JS code. Besides C and C++, Emscripten also supports other languages compiled with LLVM such as C\#, Lua, and Ruby\footnote{https://emscripten.org}. Other Wasm compilers include AssemblyScript\footnote{https://www.assemblyscript.org}, which compiles Typescript, and TeaVM\footnote{http://teavm.org}, which compiles Java bytecode. This is useful for integrating performance-sensitive code into a larger JS project or porting existing native code to the web. Wasm also allows for newer high performance languages like Rust, which programmers might prefer for its high-level ergonomics\endnotemark[\ref{en:wasm}]. As a whole, Wasm gives web developers newfound freedom in using their language of choice rather than being bound to the JS ecosystem.

In addition to compilers intended to port native code to browsers, some frameworks allow developers to build web frontends from various languages that compile to Wasm. This allows other languages to compete with JS-based tools such as Angular or React. The most popular of these is Blazor (specifically the Blazor WebAssembly variant), an open-source framework for writing single-page web application frontends in C\#\footnote{https://dotnet.microsoft.com/apps/aspnet/web-apps/blazor}. Yew is a similar alternative for writing web applications in Rust\footnote{https://github.com/yewstack/yew}.

As of 2021, Wasm-targeting web frameworks are still in their infancy, with Blazor’s production-ready official source code released in mid-2020\footnote{https://devblogs.microsoft.com/aspnet/blazor-webassembly-3-2-0-now-available/}. In addition, the use-case is still quite narrow, mainly supporting single-page, static web applications. Drawbacks such as the large payload and subsequent long loading time further suggest that, in the short-term, these frameworks will likely supplement JS-based frameworks. Despite the narrow scope, the rise of these frontend frameworks written in backend languages has provoked significant discussions on the future of client-side web development with Wasm, both for performance optimizations and simplicity of development in a single programming language.

\subsection{WebAssembly and 3D Graphics}

Since the WebXR API does not provide 3D rendering capabilities, it is used with an external WebGL library. For example, A-Frame, one of the most popular WebXR development platforms, provides full access to Three.js, which is a graphics library (\cite{franke}). Babylon.js, another popular graphics library, supports WebXR directly\footnote{https://www.babylonjs.com}. While these platforms are efficient, Wasm offers an opportunity for even further performance improvements and incorporation of native libraries into the web.

Magnum is an example of a C++ graphics engine that has utilized Wasm. Popular for its simplicity and convenience, especially for data visualization, it has recently expanded to the web by taking advantage of Emscripten\footnote{https://magnum.graphics}. By enabling efficient web demos of the engine on their website, this demonstrates how Wasm has opened the doors to web developers who might not be satisfied with JS 3D libraries or want to re-use visualizations across multiple platforms.

\subsection{WebAssembly and Biological Databases}

Implementing Wasm has also proven to be highly effective in biological database analysis and visualization by delivering near-native speeds for executing bioinformatics tools within web-based platforms (\cite{taylor}). BioWasm\footnote{https://biowasm.com}, a prime example of a biological platform leveraging Wasm, utilizes this technology to compile high-performance C/C++ genomics tools into WebAssembly, enabling these tools to run efficiently within web browsers using very little setup and without any need for local installations. By integrating Wasm, BioWasm provides a unique advantage of offering near-native performance for highly computational genomic tasks such as sequence alignment, viral molecular epidemiology analysis, and variant calling. This approach also reduces compatibility issues across different platforms and devices, creating a highly accessible, flexible, and efficient solution for running bioinformatics tools within cross-platform environments. The inherent cross-platform nature of Wasm means that tools developed using BioWasm can be run consistently on Windows, macOS, Linux, and even mobile operating systems without performance degradation or the need for platform-specific adaptations, making it a versatile and future-proof solution in the bioinformatics space.

While BioWasm has successfully implemented Wasm to run complex web-based bioinformatics computations, most biological platforms utilizing WebXR have yet to adopt this technology. Platforms such as MolecularWebXR\footnote{https://molecularwebxr.org}, which leverage WebXR for 3D visualizations of molecular structures, stand to gain significantly from incorporating Wasm into their frameworks. Integrating Wasm would drastically reduce latency for real-time visualizations, enhance bioinformatics computations through faster data processing, and enable efficient sharing of biological visualizations across devices (\cite{belghit}). Wasm’s ability to provide effective cross-platform processing means that the simulation and visualization tools on BioWasm would occur seamlessly and instantaneously, providing a more powerful and responsive experience for researchers, educators, and students alike. By enhancing collaborative scientific research, this technology would improve educational tools that aim to bring molecular biology to life and unlock new opportunities for scientific discoveries. (\cite{taghian})

\subsection{Existing WebAssembly WebXR Implementations}

While most WebXR development still does not include Wasm, there are a few frameworks and engines that have added some support. Unity has used Wasm for exporting 3D graphics to the web since 2018, but only recently added WebXR support to the WebAssembly exporter in mid-2020\footnote{https://docs.unity3d.com/Manual/index.html}. As one of the most popular XR development engines, this enabled WebXR development for its large user base without needing to learn a WebXR framework like A-Frame or Babylon that lacks many features of the Unity engine. However, Unity lacks optimization for the web, causing large file sizes and thus minimizing the performance benefits of Wasm. Similarly, LOVR\footnote{https://github.com/bjornbytes/lovr} is a Lua-based XR framework that also supports WebXR through Wasm. Finally, Wonderland Engine\footnote{https://wonderlandengine.com}, released to the public in late 2020 is a game and graphics engine dedicated to WebXR that uses Wasm, WebGL, and other performance improvements. Even though the engine is in its developmental stage, unlike Unity or other JS-based platforms, it is an excellent example of the capabilities of WebXR combined with a WebAssembly 3D engine.

In addition to development frameworks, some WebXR-adjacent libraries have begun to utilize the performance benefits of Wasm. OpenCV, a C++ computer vision library, was compiled for the web with LLVM, first using asm.js and more recently with Wasm\footnote{https://opencv.org}. Comparing it to similar JS libraries found very significant performance improvements using the Wasm version\footnote{https://hacks.mozilla.org/2017/03/why-webassembly-is-faster-than-asm-js/}\label{en:wasm_fast}. After testing out a face recognition algorithm written in three separate languages -- Wasm, asm.js, and JS -- the Wasm-powered algorithm was met with twice the frames per second (FPS) compared to its asm.js counterpart, and twenty times that compared to an equivalent JS algorithm. Considering the many uses of computer vision in AR, this example demonstrates the new XR possibilities brought by the performance benefits of Wasm.

\begin{figure}[h!]
\begin{center}
\includegraphics[width=10cm]{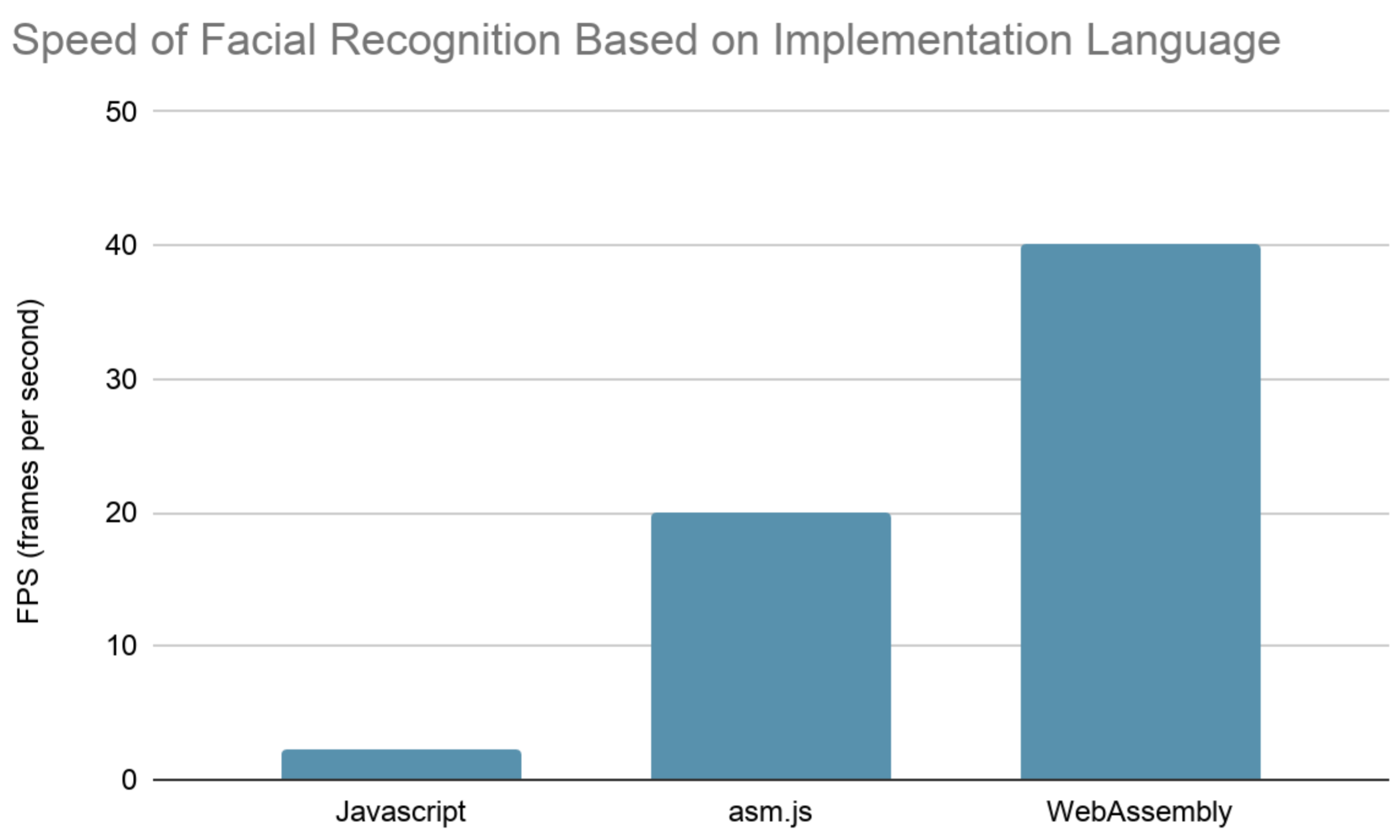}
\end{center}
\caption{FPS produced when face recognition algorithm was run with each implementation language. All the other factors, such as browser or architecture of the functionality, were kept constant during benchmarking\endnotemark[\ref{en:wasm_fast}]} \label{fig:2}
\end{figure}

\section{Discussion and Conclusions}

While WebXR remains relatively unpopular and underutilized among developers and users, improvements to performance and widespread development tools will likely bring it closer in popularity to native XR. Increased adoption and performance of WebXR has numerous benefits to the field, including economical investment that would inevitably arise with better-performing 3D games and experiences. This would facilitate further research, optimization, and development of platforms that could expand XR applications from entertainment to educational and medical fields. Studies have already highlighted its potential with big data (\cite{olshannikova}) and biological visualizations (\cite{huang}), including prototypes for complex heatmap\footnote{https://www.youtube.com/watch?v=IEDvXSbBnVU} and network\footnote{https://www.youtube.com/watch?v=Djk0sHVIMeI} graphics. With more widespread access, expensive hardware devices with exclusive walled garden software would no longer be necessary (\cite{pon}). Easy-to-use software available on the web paired with any device will most likely become the industry norm, which would additionally spur competition within that market and create an interoperable ecosystem(\cite{murphy2}). Concurrently, WebAssembly has reached a point of maturity, and more widespread integration into WebXR would further elucidate the potential for a web-based XR experience. Given that the existing infrastructure for both WebXR and Wasm is at a phase of effective usage beyond a simple proof-of-concept stage, we propose the following two recommendations/conclusions.

\subsection{Integration of WebAssembly into WebXR and other native libraries}

Given the enormous efficiency benefits of Wasm, we recommend that it be utilized within existing WebXR JS libraries to port over performance-sensitive operations, but can even stretch to the capacities of porting over entire libraries. While targeting purely expensive operations might be the most cost-effective, some have begun attempting to port entire libraries to AssemblyScript, such as glas\footnote{https://github.com/lume/glas}, which is an example of a Three.js port to AssemblyScript. Recently, there have launched many libraries that realize the benefit of WebAssembly and have similarly ported entire applications. We also believe there is a large reserve of powerful tools for native operating systems that could be utilized in WebXR using Wasm. OpenCV and Magnum, as discussed above, are relevant examples of high-performance libraries that could benefit from the WebXR domain. As a whole, we support further research and porting to improve the performance of JS libraries using this method, rather than only using Wasm to integrate existing native code into the web.

\subsection{Integration of WebXR and Wasm in Biological AR/VR Technologies}

The integration of WebXR and Wasm into biological research and education offers a transformative opportunity for advancing real-time biological database visualization and analysis (\cite{zhang}). WebXR and Wasm can be utilized to create biological platforms such as BioWasm and MolecularWebXR that allow the seamless integration and exploration of large-scale biological datasets. Combining WebXR and Wasm eliminates the need for native infrastructure and makes high-performance biological analysis more accessible across devices. Prioritizing their integration in future biological AR/VR development will promote the sharing of biological data and will enhance both research and educational outcomes in fields of genomic and proteomic data analysis. Ultimately, this technology will promote more effective collaboration by making complex biological data accessible to a wider audience. This shift will foster innovation in bioinformatics and biological database management, paving the way for more dynamic and interactive approaches for teaching and research.

\subsection{Standardization of a web-based XR framework combining WebXR and Wasm}

Similar to existing open-source solutions like A-Frame and powerful native engines like Unity, we suggest an open-source WebXR framework built specifically to focus on utilizing Wasm for its performance advantages. By focusing on accessibility, but without sacrificing complexity, such a tool could transform WebXR into the dominant form of XR experiences, and help enable XR to expand beyond entertainment purposes to other domains. Naturally, creating a standardization of a WebXR development framework is a longer-term recommendation (\cite{bondarenko}). Yet, as an increasing number of libraries that tap into XR utilize WebXR through Wasm, together with existing WebXR libraries that port operations to AssemblyScript, an industry standard is not too far on the horizon. The development, accessibility, and standardization of an open-source WebXR-Wasm framework is one that will help the XR domain reach the scalability it deserves, without relinquishing performance. The industry as a whole should head towards this exciting possibility of a write-once-deploy-anywhere workflow that can generate low latency XR applications across a multitude of use-cases, which can be fully realized with improvements to the near-universal compatibility of WebXR.


\section{Statements and Declarations}

\section*{Conflict of Interests}

 The authors declare that the research was conducted in the absence of any commercial or financial relationships that could be construed as a potential conflict of interest.

\section*{Acknowledgments}
Research reported in this publication was supported by the National Institute of Diabetes and Digestive and Kidney Diseases (NIDDK) of the National Institutes of Health (Bethesda, Maryland) (R01DK132090 to Dr Khomtchouk). The authors have reported that they have no relationships relevant to the contents of this paper to disclose.


\bibliographystyle{frontiersinSCNS_ENG_HUMS}
\bibliography{frontiers}


\end{document}